\documentstyle[preprint,aps,eqsecnum]{revtex}
\begin{document}
\draft
\preprint{UFIFT-HEP-94-4}
\title{Stretching Wiggly Strings}
\author{Jaewan Kim and Pierre Sikivie}
\address{Department of Physics, University of Florida,
Gainesville, FL 32611}
\date{\today}
\maketitle
\begin{abstract}
How does the amplitude of a wiggle on a string change when the string
is stretched? We answer this question for both longitudinal and
transverse wiggles and for arbitrary equation of state, {\it i.e.},
for arbitrary relation between the tension $\tau$ and the energy per
unit length $\epsilon$ of the string. This completes our derivation of
the renormalization of string parameters which results from averaging
out small scale wiggles on a string. The program is presented here in
its entirety.
\end{abstract}
\section{Introduction}
In this paper, we give a general treatment of the renormalization of
string parameters (the energy per unit length, the tension, and the
equation of state) when a coarse grained view of a wiggly string is
adopted. Although the formalism developed below may have applications
in other areas, our primary motivation has been to try and improve our
understanding of cosmic gauge strings\cite{vilenkin85}.
Cosmic gauge strings are a
prediction of certain grand unified theories. They may be present in
the universe today as thin line-like regions of space which are hung
up in a false vacuum of very high energy density. If one were to
discover such strings, one would have the opportunity of learning
physics in an energy regime which is  completely out of reach of
present earth-based laboratory experiments. As such, cosmic gauge
strings are worthy of careful theoretical analysis. \par
A gauge string is an example of a Nambu-Goto string,
{\it i.e.} it
has energy per unit length $\epsilon$ equal to its tension $\tau$:
$\epsilon=\tau\equiv\mu$, at the microscopic level.
On the other hand, computer simulations have shown
that cosmic gauge strings acquire small scale
structure\cite{simulation,processes}.
The effect of small scale structure is to increase the energy per unit
length of a gauge string and to decrease its tension when  a coarse
grained description of the wiggly string is
adopted\cite{carter90,vilenkin90}. Thus whereas the
``bare'' Nambu-Goto string has $\epsilon=\tau\equiv\mu$, the
renormalized wiggly Nambu-Goto (RWNG) string, {\it i.e.}
the coarse grained
description of a wiggly Nambu-Goto string, has $\bar\epsilon >\mu$
and $\bar\tau < \mu$. As a consequence, the RWNG string does not obey the
Nambu-Goto equations of motion. \par
In ref.~\cite{HKS92} were derived the
relativistic equations of motion for general strings, {\it i.e.}
strings with arbitrary relation between the tension $\tau$ and the
energy per unit length $\epsilon$, including RWNG strings.
In addition, the
renormalization of $\tau$ and $\epsilon$ that results from averaging
out small scale wiggles on a string was obtained in the general case
to lowest order in the amount of wiggliness. Let us state this latter
result more precisely as follows.\par
Consider a string with energy per unit
length $\epsilon$ and tension $\tau$. Suppose that the string
has wiggles on it with characteristic wavelength $\lambda$ but that
the string is otherwise straight. Then an observer with resolution
$d\gg\lambda$ will see a straight string, assigning to it an energy per
unit length $\bar\epsilon > \epsilon$ and tension $\bar\tau\ne\tau$.
$\bar\epsilon$ and $\bar\tau$ were calculated to lowest
order in the amount of wiggliness.
The  result has the general form:
\begin{mathletters}\label{renorm_skeleton}
\begin{eqnarray}
\bar\epsilon&=&\epsilon +
C_{\epsilon}(\epsilon,\tau,{d\tau\over d\epsilon},{d^2\tau\over d\epsilon^2},
\ldots\,)
\langle v^2\rangle \\
\bar\tau&=&\tau +
C_{\tau}(\epsilon,\tau,{d\tau\over d\epsilon},{d^2\tau\over d\epsilon^2},
\ldots\,)
\langle v^2\rangle
\end{eqnarray}
\end{mathletters}%
where $\langle v^2\rangle$ is the average (velocity)$^2$
which characterizes the wiggles. The coefficients
$C_{\epsilon}$ and $C_{\tau}$ can be found in ref.~\cite{HKS92}.
As indicated in
Eqs.~(\ref{renorm_skeleton}), these coefficients depend solely upon
the equation of state $\tau(\epsilon)$ of the original string.
The equation of state of a
string gives the relationship between $\epsilon$ and $\tau$ when the
string is slowly stretched. It must be derived from the relevant
microphysics. \par
The main purpose of the present paper is to derive the equation of
state $\bar\tau(\bar\epsilon)$ of {\it the new string}, which is the
coarse grained description of the wiggly string we started from.
To do so
we must obtain the response of $\langle v^2\rangle$ to a slow
stretching of the original wiggly string. Once the equation of state
$\bar\tau(\bar\epsilon)$ of the new string has been obtained we can
compute ${d\bar\tau\over d\bar\epsilon}$,
${d^2\bar\tau\over d{\bar\epsilon}^2}$, $\ldots\,$, and
insert these into Eqs.~(\ref{renorm_skeleton}) if we wish to average
out wiggles on the new string. The process of averaging out wiggles
may then be repeated indefinitely, moving to successively longer
scales and producing  ``running'' $\epsilon(k)$, $\tau(k)$,
${d\tau\over d\epsilon}(k)$, {\it etc}. These are the quantities
describing the string when all wiggles of wavelength shorter than
$\lambda={2\pi\over k}$ have been averaged over.\par
The plan of the paper is as follows. In section II, we obtain the
relativistic equations of motion for strings with arbitrary equation
of state. They are a generalization of the Nambu-Goto equations of
motion.
These equations were already obtained in ref.~\cite{HKS92} but we repeat the
derivation here for the sake of completeness and clarity. In section
III, we obtain the renormalization of $\epsilon$ and $\tau$ which
results from averaging out wiggles on the string. For the case of
transverse wiggles the treatment and the results are the same as in
ref.~\cite{HKS92}. For the case of longitudinal wiggles, we adopt
a new and definition of what it means to add a wiggle  to a string.
This changes
the results for the renormalization of $\epsilon$ and
$\tau$ from averaging out longitudinal wiggles compared to
those in ref.~\cite{HKS92}. We will discuss  our
motivation for adopting the new definition. In section IV, we obtain
the solution of the equations of motion which describes a
homogeneously stretching string. In section V, we study the behaviour
of small perturbations away from that solution. This will enable us to
answer the question that motivated us in the first place: how does the
amplitude of a wiggle change when the string is stretched? In section
VI, we use the results to complete our derivation of the
renormalization of string parameters from averaging out
small scale wiggles. In section VII, we apply our results to the
important special case of wiggly Nambu-Goto strings.
\section{Equations of motion for general strings}
Consider a string, {\it i.e.}, an object whose stress-energy-momentum
is localized on a line in space. Let
$X^{\mu}(\sigma)$ be the space-time coordinates of the worldsheet
with respect to a Lorentz reference frame.
$\sigma = (\sigma^0,\sigma^1)$ are arbitrarily chosen coordinates
parametrizing points on the worldsheet.
The metric induced on the worldsheet is
\begin{equation}
h_{ab} (\sigma) =
\partial_a X^\mu \partial_b X^\nu g_{\mu\nu}(x),
\qquad a,b=0,1\,,
\label{metric}
\end{equation}
where $g_{\mu\nu}$ is the space-time metric. For the purpose of this paper
we will restrict ourselves to the case
$g_{\mu\nu}=\eta_{\mu\nu}=\mbox{diag}(1,-1,-1,-1)$.
At point $X^\mu(\sigma)$ on the string lives the
$2$-dim.~stress-energy-momentum tensor
\begin{equation}
t^{ab} = (\epsilon - \tau) u^a u^b + \tau\,h^{ab}, \label{emtensor}
\end{equation}
where $\epsilon(\sigma)$ is the energy per unit length of the
string, $\tau (\sigma)$ is its tension and $u^a (\sigma)$ is the
fluid velocity parallel to the string: $u^a u_a = +1$.  The
$4$-dim.~stress-energy-momentum tensor is then
\begin{equation}
T^{\mu\nu}(x) = \int d^2\sigma \sqrt{-h}\,
t^{ab} (\sigma) \partial_a X^\mu (\sigma) \partial_b X^\nu (\sigma) \delta^4
\Bigl( x-X (\sigma)\Bigr)\ , \label{4emtensor}
\end{equation}
where $h = \det(h_{ab})$.
This expression is both invariant under $2$-dim.~reparametrizations
and covariant under $4$-dim.~Lorentz transformations.
The motion of the string must conserve $4$-dim.~energy and
momentum. Thus
\begin{eqnarray}
0&=&\partial_{\mu}T^{\mu\nu}(x)=\int d^2\sigma \sqrt{-h}\,
t^{ab}\partial_a X^\mu\partial_b X^\nu
\partial_{\mu}\delta^4 \Bigl( x-X (\sigma)\Bigr)    \nonumber \\
&=&\int d^2\sigma \sqrt{-h}\,
t^{ab}\partial_b X^\nu\partial_a
\delta^4 \Bigl( x-X (\sigma)\Bigr) \nonumber \\
&=&-\int d^2\sigma \,\partial_a \Bigl[\sqrt{-h}
t^{ab}\partial_b X^\nu \Bigr]
\delta^4 \Bigl( x-X (\sigma)\Bigr) .
\end{eqnarray}
The equations of motion for general strings are therefore
\begin{equation}
\partial_a \Bigl[ \sqrt{-h}\, t^{ab} (\sigma)
\partial_b X^\mu \Bigr] = 0\quad(\mu=0,1,2,3)\ . \label{4eqmo}
\end{equation}
To complete the description of the string dynamics, we must
supplement Eq.~(\ref{4eqmo})
with an equation of state:
\begin{equation}
\tau = \tau (\epsilon)\ . \label{eqst}
\end{equation}
We then have five equations for the five unknowns:
$\vec X_\perp (\sigma)$, $\epsilon (\sigma)$, $\tau (\sigma)$, and
$\beta(\sigma)\equiv{u^1 (\sigma)\over u^0 (\sigma)}$.
$\vec X_\perp (\sigma)$
represents the two transverse degrees of freedom of the string.
$\beta (\sigma)$ is its longitudinal velocity.  The five equations
uniquely specify the time evolution that results from
arbitrary initial conditions.
It was shown in ref.~\cite{HKS92}
that the case of the Nambu-Goto string is included in this description.
It corresponds to the equation of state $\epsilon=\tau$.\par
Eqs.~(\ref{4eqmo}) and (\ref{eqst}) describe how the fluid
motion along the string affects the motion of the string in the
transverse directions, and vice versa. This dynamics is purely
classical. However, there are also interesting quantum mechanical
effects associated with the addition of new degrees of freedom
to a string. Indeed,  consider a straight
string at rest. From a classical viewpoint, there are in this case no
forces of the string upon the fluid attached to it. However, from a
quantum mechanical viewpoint, there are forces of the string
upon the fluid because
the fluid distribution affects the quantum mechanical fluctuations of
the string. In ref.~\cite{ds93}, these forces were calculated for the
case of beads attached to a straight string. In the present paper, such
order $\hbar$ effects are neglected.
\section{Renormalization of $\epsilon$ and $\tau$ due to wiggliness}
In this section, we derive in detail the renormalization of the
energy per unit length and the tension of a wiggly string
which must be performed when a coarse grained
description of the string is adopted.
Consider first a straight motionless string lying along the $x$-axis.
This is a trivial static solution to Eqs.~(\ref{4eqmo}). It is
given in the $\sigma=(t,x)$ gauge by
\begin{mathletters}\label{stat-sol}
\begin{eqnarray}
X^{\mu}(\sigma)&=&(t,x,0,0)  \\
\epsilon(t,x)&=&\epsilon_0=\mbox{constant} \\
\tau(t,x)&=&\tau(\epsilon_0)=\tau_0  \\
\beta(t,x)&=&0.
\end{eqnarray}
\end{mathletters}%
Next we perturb the string,  exciting small oscillations about the
static solution. In a coarse grained view, the string appears straight and
motionless again, but with the $4$-dim.~stress-energy-momentum
tensor:
\begin{equation}
\overline{T}^{\mu\nu}=\mbox{diag}(\bar\epsilon,-\bar\tau,0,0)
\delta(y)\delta(z)=\langle T^{\mu\nu}\rangle, \label{new-tensor}
\end{equation}
where $T^{\mu\nu}$ is the stress-energy-momentum tensor of the wiggly
string 
and $\langle T^{\mu\nu}\rangle$ is the space-time average thereof.
We have assumed, as we do throughout this paper, that the wiggles do not
carry average momentum with respect to the Lorentz frame in which the
unperturbed string is at rest.
In order to compute  $\langle T^{\mu\nu}\rangle$,
it is necessary to solve the linearized equations of motion for
the wiggles with appropriate initial conditions.\par
What are the appropriate initial conditions?
Equivalently, one might ask: what does it mean to add a wiggle to
a string? Throughout this paper we will adopt the following
definition. Consider a length $L$ of string whose unperturbed state is
given by Eqs.~(\ref{stat-sol}). At time $t=0$, apply sudden infinite
(impulse type) forces to the string which instantaneously change its
local longitudinal and transverse velocities without changing any
positions. After $t=0$, let the string move freely.
It can be shown that,
for the purpose of the present paper, that definition
is equivalent to the following one. Consider a length $L$ of string whose
unperturbed state is given by Eq.~(\ref{stat-sol}). Stretch this string
longitudinally and/or transversely keeping the position of the end points
fixed
in the laboratory frame of reference. Then let the string move freely.\par
For transverse
wiggles in the $y$-direction, the initial condition consistent with
our definition is
\begin{mathletters}\label{trans-init-cond}
\begin{eqnarray}
X^{\mu}(0,x)&=&(0,x,0,0)  \\
\partial_t X^{\mu}(0,x)&=& (1,0,\dot{y}_{\rm in}(x),0) \\
\beta(0,x)&=&0 \\
\epsilon(0,x)&=&\epsilon_0 \\
\tau(0,x)&=&\tau_0,
\end{eqnarray}
\end{mathletters}%
where $\dot{y}_{\rm in}(x)$ is the initial transverse velocity at
point $x$. Eqs.~(\ref{trans-init-cond}) are the initial conditions for
transverse wiggles used in ref.~\cite{HKS92}. For longitudinal wiggles, the
initial condition is
\begin{mathletters}\label{long-init-cond}
\begin{eqnarray}
X^{\mu}(0,x)&=&(0,x,0,0)  \\
\partial_t X^{\mu}(0,x)&=& (1,0,0,0) \\
\beta(0,x)&=&\beta_{\rm in}(x) \\
\epsilon(0,x)&=&\epsilon_0 -{1\over 2}(\epsilon_0-\tau_0)
{\beta_{\rm in}}^2(x)+{\cal O}({\beta_{\rm in}}^4) \label{licd}\\
\tau(0,x)&=&\tau(\epsilon(0,x))\label{lice},
\end{eqnarray}
\end{mathletters}%
where $\beta_{\rm in}(x)$ is the initial longitudinal velocity at
point $x$. Eqs.~(\ref{long-init-cond}) are not the initial conditions
for longitudinal wiggles used in ref.~\cite{HKS92}.
In ref.~\cite{HKS92}, we set
$\epsilon(0,x)=\epsilon_0$ and $\tau(0,x)=\tau_0$.
We will give the reasons for the change in section V. With the
definition adopted in this paper of what it means to add a wiggle to a
string, we must use Eqs.~(\ref{long-init-cond}).
Indeed, consider a small length $\Delta L$ of string at point $x$. At
time $t=0$, since positions do not change instantaneously, that length
does not change in the laboratory frame of reference. Therefore,
because of Lorentz contraction, in the local instantaneous rest frame
of the string after $t=0$, that small piece of string has length
\begin{equation}
\Delta L[1-{\beta_{\rm in}}^2(x)]^{-1/2}=
\Delta L[1+{1\over 2}{\beta_{\rm in}}^2(x)]+
{\cal O}({\beta_{\rm in}}^4).
\end{equation}
In its rest frame, the piece of string got stretched.
The stretching of the proper length
$\Delta L$ by the amount
$d\Delta L = {1\over 2}{\beta_{\rm in}}^2(x) \Delta L$
produces a decrease $d\epsilon$ in the energy per unit length which is
readily obtained by equating the change in energy with the work done:
\begin{equation}
d(\epsilon\Delta L)=\tau d\Delta L
\end{equation}
which yields
\begin{equation}
d\epsilon=-(\epsilon-\tau){d\Delta L\over \Delta L}=
 -{1\over 2}(\epsilon-\tau)\beta^2_{\rm in},
\end{equation}
which in turn yields Eq.~(\ref{licd}). Eq.~(\ref{lice}) then follows
from the equation of state.\par
We are now ready to proceed. Let us first analyze longitudinal
wiggles. The longitudinally perturbed string is described in the
$\sigma=(t,x)$ gauge by
\begin{mathletters}\label{}
\begin{eqnarray}
X^{\mu}(t,x)&=&(t,x,0,0)  \\
\beta(t,x)&=&\beta_1(t,x)+\beta_2(t,x)+\ldots\, \\
\epsilon(t,x)&=&\epsilon_0+\epsilon_1(t,x)+\epsilon_2(t,x)+\ldots\, \\
\tau(t,x)&=&\tau_0+\tau_1(t,x)+\tau_2(t,x)+\ldots\,\, ,
\end{eqnarray}
\end{mathletters}%
where the expansions are in powers of $\beta_{\rm in}$. The
equation of state relates $\tau$ and $\epsilon$. Hence
\begin{mathletters}\label{higher-taus}
\begin{eqnarray}
\tau_1&=& +{d\tau\over d\epsilon}\Big |_0\epsilon_1
=-v_{\scriptscriptstyle L,0}^2\epsilon_1, \\
\tau_2&=&-v_{\scriptscriptstyle L,0}^2\epsilon_2
+{1\over 2}{d^2\tau\over d\epsilon^2}\Big|_0\epsilon_1^2,
\end{eqnarray}
\end{mathletters}%
where $v_{\scriptscriptstyle L,0}\equiv\sqrt{-{d\tau\over d\epsilon}|_0}$.
The $2$-dim.~stress-energy-momentum tensor is
\begin{equation}
\Bigl(t^{ab}\Bigr)=
\pmatrix{(\epsilon-\tau)\gamma^2+\tau &
(\epsilon-\tau)\gamma^2\beta \cr
(\epsilon-\tau)\gamma^2\beta &
(\epsilon-\tau)\gamma^2\beta^2-\tau },
\end{equation}
where $\gamma=(1-\beta^2)^{-1/2}$. The linearized equations of motion are
\begin{mathletters}\label{lin-eq}
\begin{eqnarray}
\dot{\epsilon_1}+(\epsilon_0-\tau_0)\beta_1'&=&0 \\
(\epsilon_0-\tau_0)\dot{\beta}_1+
v_{\scriptscriptstyle L,0}^2\epsilon_1'&=&0,
\end{eqnarray}
\end{mathletters}%
where dots and primes denote derivatives with respective to $t$ and
$x$ respectively. If the longitudinal oscillation is initiated by
giving longitudinal velocity
$\beta_{\rm in}(x)=\beta_{\rm in}\sin{kx}$ at time $t=0$, then
the solution to Eqs.~(\ref{lin-eq}) and (\ref{long-init-cond}) is
\begin{mathletters}\label{linsol}
\begin{eqnarray}
\beta_1(t,x)&=&\beta_{\rm in}\sin{kx}\,
\cos{v_{\scriptscriptstyle L,0}kt} \label{lsa} \\
\epsilon_1(t,x)&=&
-(\epsilon_0-\tau_0){\beta_{\rm in}\over v_{\scriptscriptstyle L,0}}
\cos{kx}\,\sin{v_{\scriptscriptstyle L,0}kt}.\label{lsb}
\end{eqnarray}
\end{mathletters}%
Eqs.~(\ref{linsol}) show that, as was implied by our notation,
$v_{\scriptscriptstyle L,0}$
is the phase velocity of longitudinal wiggles.
$\langle T^{\mu\nu}\rangle$ must be computed up to
second order in $\beta_{\rm in}$. Using Eqs.~(\ref{4emtensor}),
(\ref{new-tensor}), (\ref{long-init-cond}), (\ref{higher-taus}),
and the fact that
$\langle\epsilon_1\rangle=\langle\beta_1\rangle=0$,  we find
\begin{mathletters}
\label{sec-emtensor}
\begin{eqnarray}
\bar\epsilon=\langle t^{00}\rangle&=&
      \epsilon_0+\langle\epsilon_2\rangle+
      (\epsilon_0-\tau_0)\langle\beta_1^2\rangle \\
0=\langle t^{01}\rangle &=&
      (\epsilon_0-\tau_0)\langle\beta_2\rangle+
      (1+v_{\scriptscriptstyle L,0}^2)\langle\epsilon_1\beta_1\rangle
      \label{t01}  \\
-\bar\tau=\langle t^{11}\rangle&=&
      -\tau_0+(\epsilon_0-\tau_0)\langle\beta_1^2\rangle+
      v_{\scriptscriptstyle L,0}^2\langle\epsilon_2\rangle-
      {1\over 2}{d^2\tau\over d\epsilon^2}\Big |_0
      \langle\epsilon_1^2\rangle.
\end{eqnarray}
\end{mathletters}%
Eq.~(\ref{t01}) expresses our assumption that there is no net flow of
momentum in either direction.
To determine $\langle\epsilon_2\rangle$ we use the equation of motion
in the time-like direction, to second order in $\beta_{\rm in}$:
\begin{equation}
\dot{\epsilon_2}+2(\epsilon_0-\tau_0)\beta_1\dot{\beta_1}+
(\epsilon_0-\tau_0)\beta_2'+
(1+v_{\scriptscriptstyle L,0}^2)[\epsilon_1\beta_1]'=0 .
\label{sec-eqmo}
\end{equation}
Averaging Eq.~(\ref{sec-eqmo}) over an interval $0\le x\le L$, we obtain
\begin{equation}
\partial_t{1\over L}\int_0^L\,
dx[\epsilon_2+(\epsilon_0-\tau_0)\beta_1^2] ={\cal O}({1\over L}).
\end{equation}
This equation merely tells us that the average energy associated with
the longitudinal perturbation is time-independent. So we may evaluate
the average energy from the initial condition alone, using
Eqs.~(\ref{licd}) and (\ref{lsa}):
\begin{eqnarray}
\langle\epsilon_2\rangle+(\epsilon_0-\tau_0)\langle\beta_1^2\rangle
&=&{1\over L}\int_0^L dx
\Big[\epsilon_2+(\epsilon_0-\tau_0)\beta_1^2\Big]_{t=0} \nonumber\\
&=&{1\over 4}(\epsilon_0-\tau_0)\beta_{\rm in}^2
=(\epsilon_0-\tau_0)\langle\beta_1^2\rangle . \label{epsi1}
\end{eqnarray}
Hence
\begin{equation}
\langle\epsilon_2\rangle=0. \label{epsi2}
\end{equation}
The renormalized energy per unit length and tension due to
longitudinal perturbations are therefore
\begin{mathletters}\label{long-renorm}
\begin{eqnarray}
\bar\epsilon_{\scriptscriptstyle L}
&=&\epsilon_0+(\epsilon_0-\tau_0)\langle\beta_1^2\rangle \\
\bar\tau_{\scriptscriptstyle L}
&=&\tau_0-(\epsilon_0-\tau_0)\langle\beta_1^2\rangle
 \Big[1+ (\epsilon_0-\tau_0)
 {d\ln{v_{\scriptscriptstyle L}}\over d\epsilon}\Big|_0\Big],
\end{eqnarray}
\end{mathletters}%
to lowest non-trivial order in the amount of wiggliness. The amount of
longitudinal wiggliness is characterized in our formalism by
$\langle\beta_1^2\rangle$. Eqs.~(\ref{long-renorm}) differ from the
corresponding equations in ref.~\cite{HKS92} because we use here a different
definition of what is meant by adding a wiggle to the string. The
inadequacy of the definition used in ref.~\cite{HKS92} did not become apparent
to us until we studied the behaviour of the wiggles when the string is
being stretched (see section V).\par

Next let us consider transverse wiggles. In the gauge
$\sigma =(t,x)$,  the space-time coordinates of a wiggly string whose
average direction is along the $x$-axis is
\begin{equation}
(X^\mu(t,x))=(t,x,y(t,x),z(t,x)),
\end{equation}
where $y(t,x)$ and $z(t,x)$ average to zero.
The induced metric of the worldsheet is
\begin{equation}
(h_{ab})=\pmatrix{1-\dot{y}^2-\dot{z}^2 & -\dot{y}y'-\dot{z}z' \cr
                         -\dot{y}y'-\dot{z}z'& -1-{y'}^2-{z'}^2}.
\end{equation}
Let us write
\begin{mathletters}
\begin{eqnarray}
\epsilon(t,x)&=&\epsilon_0+\epsilon_1(t,x) \\
\tau(t,x)&=&\tau_0-v_{\scriptscriptstyle L,0}^2
   \epsilon_1(t,x)+{\cal O}(\epsilon_1^2).
\end{eqnarray}
\end{mathletters}%
We will see below that $\epsilon_1$ and $\beta$ are of second order
in $y$ and $z$. The $2$-dim.~stress-energy-momentum tensor up to
second order in $y$ and $z$ is then
\smallskip
\begin{equation}
(t^{ab})=
\pmatrix{\epsilon_0+\epsilon_1+\epsilon_0 (\dot{y}^2+\dot{z}^2)&
             (\epsilon_0-\tau_0)\beta-\tau_0(\dot{y}y'+\dot{z}z') \cr
             (\epsilon_0-\tau_0)\beta-\tau_0(\dot{y}y'+\dot{z}z') &
              -\tau_0+v_{\scriptscriptstyle L,0}^2\epsilon_1+
              \tau_0({y'}^2+{z'}^2)}.\label{trans-emtensor}
\end{equation}
\smallskip
Because of rotational symmetry around the $x$-axis and the lack of
mixing between the $y$- and $z$-components of the wiggles
up to second order, we set $z=0$ and
focus solely on the $y$-component from  hereon.
Up to second order in $y$, the equations of motion are
\begin{mathletters}\label{trans-eq-mo}
\begin{eqnarray}
&&\epsilon_0\ddot{y}-\tau_0{y''}=0  \label{tema}\\
&&\dot{\epsilon_1}+(\epsilon_0-\tau_0)\beta'+
(\epsilon_0-\tau_0)\dot{y}'y'=0 \label{temb}\\
&&(\epsilon_0-\tau_0)\dot\beta+
v_{\scriptscriptstyle L,0}^2\epsilon_1'+
\tau_0y'(y''-\ddot{y})=0.\label{temc}
\end{eqnarray}
\end{mathletters}%
Eq.~(\ref{tema}) tells us that transverse wiggles are sinusoidal
waves with phase velocity
$v_{\scriptscriptstyle T,0}=\sqrt{\tau_0\over\epsilon_0}$.
Eqs.~(\ref{temb}) and (\ref{temc}) may be combined to produce
wave equations for $\epsilon_1$
and  $\beta$ with source terms:
\begin{mathletters}\label{wave-eq}
\begin{eqnarray}
\ddot\epsilon_1-v_{\scriptscriptstyle L,0}^2\epsilon_1''&=&
-{1\over 2}[\epsilon_0\partial_t^2-\tau_0\partial_x^2]
(\dot{y}^2+{y'}^2) \label{wea}\\
\ddot\beta-v_{\scriptscriptstyle L,0}^2\beta''&=&
{v_{\scriptscriptstyle L,0}^2-v_{\scriptscriptstyle T,0}^2
\over 2(1-v_{\scriptscriptstyle T,0}^2)}
\partial_t\partial_x(\dot{y}^2+{y'}^2)+
{v_{\scriptscriptstyle T,0}^2\over 1-v_{\scriptscriptstyle T,0}^2}
[\partial_t^2-v_{\scriptscriptstyle L,0}^2\partial_x^2]\dot{y}y'.
\label{web}
\end{eqnarray}
\end{mathletters}%
These equations show that transverse wiggles induce perturbations in
$\epsilon$ and $\beta$ which are second order in $y$.\par
Let us initiate the transverse wiggle by imposing the initial
conditions (\ref{trans-init-cond}) with
$\dot{y}_{\rm in}(x)=\dot{y}(0,x)=
y_{\rm in}v_{\scriptscriptstyle T,0}k\sin{kx}$. Then  Eq.~(\ref{tema})
implies
\begin{equation}
y(t,x)=y_{\rm in}v_{\scriptscriptstyle T,0}k\sin{kx}
\sin{v_{\scriptscriptstyle T,0}kt}.\label{y-sol}
\end{equation}
{}From Eq.~(\ref{temb}), we obtain
\begin{equation}
\partial_t{1\over L}\int_0^L dx
[\epsilon_1+{1\over 2}(\epsilon_0-\tau_0){y'}^2]=
{\cal O}({1\over L}).
\end{equation}
Hence
\begin{eqnarray}
&&\langle\epsilon_1\rangle+{1\over 2}(\epsilon_0-\tau_0)
\langle y'^2\rangle  \nonumber \\
&=&{1\over L}\int_0^L dx
\Big[\epsilon_1+{1\over 2}(\epsilon_0-\tau_0){y'}^2)\Big]_{t=0}=0,
\end{eqnarray}
since $y'(0,x)=\epsilon_1(0,x)=0$.
Therefore,
\begin{eqnarray}
\langle\epsilon_1\rangle&=&
-{1\over 2}(\epsilon_0-\tau_0)\langle {y'}^2\rangle \nonumber\\
&=&-{1\over 2}(\epsilon_0-\tau_0){\epsilon_0\over\tau_0}
\langle \dot{y}^2\rangle  \label{trans-epsi1on1-average}
\end{eqnarray}
{}From Eqs.~(\ref{4emtensor}), (\ref{new-tensor}),
(\ref{trans-emtensor}) and (\ref{trans-epsi1on1-average})
we obtain
\begin{mathletters}\label{trans-renorm}
\begin{eqnarray}
\bar\epsilon_{\scriptscriptstyle T}
       &=&\langle\sqrt{-h}t^{00}\rangle=\epsilon_0+\epsilon_0
       (\langle \dot{y}^2\rangle+\langle \dot{z}^2\rangle) \\
\bar\tau_{\scriptscriptstyle T}
       &=&-\langle\sqrt{-h}t^{11}\rangle=
       \tau_0-{1\over 2}
       (\langle \dot{y}^2\rangle+\langle \dot{z}^2\rangle)
       [\tau_0+\epsilon_0
       +v_{\scriptscriptstyle L,0}^2\epsilon_0
       (1-{\epsilon_0\over\tau_0})],
\end{eqnarray}
\end{mathletters}%
for the renormalized energy per unit length and tension from averaging
out small scale transverse wiggles.
Combining Eqs.~(\ref{long-renorm}) and (\ref{trans-renorm}),  the
renormalization group equations (RGE)
for $\epsilon(k)$ and $\tau(k)$ are
\begin{mathletters}\label{RGE}
\begin{eqnarray}
-{d\epsilon\over d\ln k}&=&W_T(k)\epsilon+W_L(k)(\epsilon-\tau) \\
-{d\tau\over d\ln k}&=&-{1\over 2}W_T(k)
\Big[\tau+\epsilon+v_{\scriptscriptstyle L}^2\epsilon
(1-{\epsilon\over\tau})\Big]
\nonumber \\
&&\quad -W_L(k)(\epsilon-\tau)
\Big[1+
(\epsilon-\tau){d\ln v_{\scriptscriptstyle L}\over d\epsilon}\Big],
\end{eqnarray}
\end{mathletters}%
where $W_T(k)$ and $W_L(k)$ are the spectral densities on a $\ln k$
scale of  $\langle\dot{y}^2\rangle +\langle\dot{z}^2\rangle$ and
$\langle\beta^2\rangle$ respectively.\par
However, as was already emphasized in ref.~\cite{HKS92} and also in the
introduction, Eqs.~(\ref{RGE}) are not by themselves complete
since the equation of state, as given by
$v_{\scriptscriptstyle L}^2=-{d\tau\over d\epsilon}$,
${d\ln{v_{\scriptscriptstyle L}}\over d\epsilon}=
-{1\over 2v_{\scriptscriptstyle L}^2}
{d^2\tau\over d\epsilon^2}$, {\it etc.}, is also modified by the
wiggles and hence varies from scale to scale. To obtain the
modification of the equation of state from averaging out small scale
wiggles we must study the response of the wiggles to an adiabatic
stretching of the string.
\section{homogeneous stretching solutions}
In this section we find the solution of the equations of motion which
describes a straight string which is stretching uniformly. Later we will
add wiggles onto this background solution and we will study how the
wiggles' amplitudes respond to the adiabatic stretching. \par
Our homogeneous stretching ansatz is that the gradient of $\epsilon$,
and hence the gradient of $\tau$ as well, be everywhere
parallel to the local $2$-velocity:
\begin{equation}
\partial_a\epsilon=C(\sigma)u_a.\label{ansatz}
\end{equation}
Using $u^au_a=1$, one readily obtains
\begin{equation}
(h^{ab}-u^a u^b)\,\partial_b\epsilon=0, \label{same-ansatz}
\end{equation}
as an equivalent expression.\par
Let us assume that the string is lying along the $x$-axis.
In the $\sigma=(t,x)$ gauge, Eq.~(\ref{ansatz}) is equivalent to
\begin{equation}
\partial_x\epsilon+\beta\partial_t\epsilon=0. \label{ansatz-lab-gauge}
\end{equation}
Also in this gauge the equations of motion (\ref{4eqmo}) become simply
\begin{equation}
\partial_a  t^{ab}=0, \quad  b=0,1\, , \label{lab-eqmo}
\end{equation}
where $t^{ab}$ is given by Eq.~(\ref{emtensor}) with
$(h^{ab})=(\eta^{ab})=\mbox{diag}(1,-1)$.\par
The component of Eqs.~(\ref{lab-eqmo}) which is perpendicular to
$u^a$ [{\it i.e.}, $(h^c_b-u^c u_b)\partial_a\,t^{ab}=0$] and
Eq.~(\ref{same-ansatz}) together imply
\begin{equation}
u^c\partial_c u^b=0, \label{geodesic}
\end{equation}
which is equivalent to the statement that the trajectory of any
comoving point on the string has zero acceleration:
\begin{equation}
{d\beta\over dt}=\partial_t\beta+\beta\partial_x\beta=0.
\label{same_geodesic}
\end{equation}
This result can be easily understood.
Indeed,  because the string is homogeneously
stretching, in the instantaneous rest frame of any physical point of
the string, the string on both sides of the point appears the same
({\it i.e.}, the right hand side is the reflection of the left hand
side) and hence there can be no force on the point by symmetry.\par
Consider the physical point which at time $t=0$ is located at
$x=x_0$. At later times, it is located at
\begin{equation}
x(t,x_0)=x_0+t\beta_0(x_0), \label{position}
\end{equation}
where $\beta_0(x_0)=\beta(0,x_0)$. We have
\begin{equation}
\beta(t,x(t,x_0))=\beta_0(x_0).\label{velocity}
\end{equation}
Taking the derivative of Eq.~(\ref{velocity}) with respect to $x_0$,
we obtain
\begin{equation}
\partial_x\beta(t,x(t,x_0))={H(x_0)\over 1+t\,H(x_0)},
\label{d-x-beta}
\end{equation}
where $H(x_0)\equiv{d\beta_0\over dx_0}$.\par
Next, let us analyze the component of the equations of motion which is
parallel to $u^a$: $u^b\partial_a\,t^{ab}=0$.
This may be rewritten as
\begin{equation}
u^{a}\partial_{a}\epsilon = -(\epsilon-\tau)\partial_{a}u^{a},
\end{equation}
and then again as
\begin{equation}
{d\epsilon\over dt}=\partial_{t}\epsilon+\beta\partial_{x}\epsilon=
-(\epsilon-\tau)\partial_{x}\beta.\label{d-epsilon-d-t}
\end{equation}
We may integrate Eq.~(\ref{d-epsilon-d-t})
along the trajectory of any physical
point on the string. Using Eq.~(\ref{d-x-beta}), this yields
\begin{equation}
\int_{\epsilon(t_1,x(t_1,x_0))}^{\epsilon(0,x_0)}
  {d\epsilon\over\epsilon-\tau(\epsilon)}
= \int_0^{t_1}{H(x_0)\over 1+t\, H(x_0)}dt
= \ln\Big[1+t_1\, H(x_0)\Big]. \label{integral-trajectory}
\end{equation}
Eq.~(\ref{integral-trajectory})
determines $\epsilon(t,x)$ in terms of the initial conditions
$\epsilon(0,x_0)$ and $\beta(0,x_0) \equiv \beta_0(x_0)$.
We now demand that the resulting $\epsilon(t,x)$ satisfy
the homogeneous stretching ansatz (\ref{ansatz-lab-gauge}).
Taking the derivative of
Eq.~(\ref{integral-trajectory}) with respect to
$x_0$ and using Eq.~(\ref{ansatz-lab-gauge}), we obtain
\begin{equation}
{d\epsilon(0,x_0)\over dx_0}=(\epsilon(x_0)-\tau(x_0))
\Big[{\beta_0(x_0)H(x_0)\over 1-\beta_0(x_0)^2} +
{t_1{dH\over dx_0}\over 1+t_1\, H(x_0)}\Big].
\end{equation}
This requires ${dH\over dx_0} = 0$ or
\begin{equation}
\beta_0(x_0)=Hx_0, \quad \mbox{where}\quad
H = {\mbox{constant}}, \label{beta_0}
\end{equation}
and
\begin{equation}
{d\epsilon(0,x_0)\over dx_0}=\Big[\epsilon(x_0)-\tau(x_0)\Big]
{Hx_0\over 1-H^2x^2_0}.\label{epsilon-init-eq}
\end{equation}
Eqs.~(\ref{beta_0}) and (\ref {epsilon-init-eq})
give the initial conditions which
produce a homogeneous stretching solution. Note that the solution
depends on only two parameters:
$H$ and $\epsilon(0,0)$. Note also that the solution exists only in a
limited region of space-time. At the initial time $(t=0)$, the
solution exists only for $\mid x_0\mid < H^{-1}$. At other times,
the solution exist only for $\mid x \mid < H^{-1} + t$. See
Fig.~I.\par
This homogeneous stretching solution may be obtained in
another, possibly more elegant, way. Let us choose coordinates
$(\lambda,\sigma)$ such that a comoving point on the string is always
labeled by the same space-like coordinate $\sigma$, and such that the
direction of the time-like coordinate $\lambda$ is everywhere
orthogonal to that of $\sigma$, in the Minkowskian sense. In this
coordinate system, $u^1=0$ and hence the homogeneity condition
(\ref{ansatz}) implies $\epsilon = \epsilon(\lambda)$. The induced metric
is still diagonal however. Since $\epsilon$ does not depend upon
$\sigma$, we may assume, as part of our homogeneity ansatz, that the
metric also does not depend upon $\sigma$. We may then choose
$\lambda$ in such a way that
\begin{equation}
(h_{ab}) = a^2(\lambda)\pmatrix{1&0\cr 0&-1},
\end{equation}
which is equivalent to the following three conditions relating $(t,x)$
and $(\lambda,\sigma)$:
\begin{mathletters}\label{metric-comp}
\begin{eqnarray}
&&\dot{t}^2-\dot{x}^2= a^2(\lambda)\\
&&\dot{t}t'-\dot{x}x'=0\\
&&\dot{t'}^2-\dot{x'}^2=-a^2(\lambda)
\end{eqnarray}
\end{mathletters}%
where the dots and primes denote derivatives with respect to $\lambda$
and $\sigma$ respectively. Eqs.~(\ref{metric-comp}) may be rewritten as
\begin{equation}
  {\partial(t,x)\over\partial(\lambda,\sigma)}=a(\lambda)
  \pmatrix{\cosh\psi(\lambda,\sigma)&\sinh\psi(\lambda,\sigma)\cr
    \sinh\psi(\lambda,\sigma)&\cosh\psi(\lambda,\sigma)},\label{jacobian}
\end{equation}
where $\psi(\lambda,\sigma)$ and $a(\lambda)$ are as yet unknown
functions. They are constrained by the commutativity of
partial derivatives which implies
\begin{mathletters}
\begin{eqnarray}
a\psi'\sinh\psi&=&\dot{a}\sinh\psi+a\dot{\psi}\cosh\psi \\
a\psi'\cosh\psi&=&\dot{a}\cosh\psi+a\dot{\psi}\sinh\psi.
\end{eqnarray}
\end{mathletters}%
These equations in turn imply
\begin{mathletters}
\begin{eqnarray}
(a\psi' - \dot{a})^2&=&a^2\dot{\psi}^2 \\
a\dot{\psi}\tanh\psi &=& a\psi' -\dot{a},
\end{eqnarray}
\end{mathletters}%
which in turn imply
\begin{equation}
\psi'-{\dot{a}\over a} = \dot{\psi}\tanh\psi=\pm \dot{\psi}.
\end{equation}
The solution $\psi=\mbox{constant}$, $a=\mbox{constant}$,
with $\tanh\psi=\pm 1$,
does not yield an invertible metric and hence is not acceptable.
The unique remaining solution is
\begin{equation}
\psi = H \sigma , \quad a=e^{H\lambda}
\end{equation}
where $H$ is a constant. Substituting this back into
Eqs.~(\ref{jacobian})
and solving the corresponding partial differential equations for
$t(\lambda,\sigma)$ and $x(\lambda,\sigma)$, we obtain
\begin{mathletters}\label{transformation}
\begin{eqnarray}
t&=&{1\over H}(e^{H\lambda}\cosh H\sigma -1) \\
x&=&{1\over H}e^{H\lambda}\sinh H\sigma,
\end{eqnarray}
\end{mathletters}%
where the constants of integration were fixed by requiring the
point $(t,x)=(0,0)$ to be mapped onto the point
$(\lambda,\sigma)=(0,0)$. See Fig.~I.\par
The $2$-velocity in the $(\lambda,\sigma)$ coordinates is
$({1\over a},0)$. By transforming back to the $(t,x)$ coordinates, we
find
\begin{equation}
\beta(t,x)={u^1\over u^0}\Big|_{(t,x)}=\tanh H\sigma
 ={Hx\over 1+Ht},
\end{equation}
which is consistent with our previous description
[{\it cf.} Eqs.~(\ref{d-x-beta}) and (\ref{beta_0})].
The only independent equation of motion in $(\lambda,\sigma)$
coordinates is
\begin{equation}
\dot{\epsilon} = -(\epsilon - \tau)H. \label{eq-epsilon_0}
\end{equation}
Again, this is consistent with our previous description
[{\it cf.} Eqs.~(\ref{d-epsilon-d-t}) and
(\ref{epsilon-init-eq})].
Eq.~(\ref{eq-epsilon_0}) determines $\epsilon(\lambda)$
for a given equation of
state. For example, the renormalized wiggly Nambu-Goto string equation
of state, $\epsilon\tau=\mu^2$ (see section VII), yields
\begin{equation}
\epsilon(\lambda)=\sqrt{\mu^2+e^{-2H\lambda}(\epsilon_i^2-\mu^2)},
\end{equation}
where $\epsilon_i$ is the energy per unit length on the $\lambda=0$
hypersurface: $(1+Ht)^2-H^2x^2=1$.
\section{stretching a wiggly string}
We now add wiggles as small deviations away from the solution
describing a homogeneously stretching string. In the comoving
orthogonal gauge, $(\sigma^0,\sigma^1)\equiv(\lambda,\sigma)$,
introduced in the previous section, the unperturbed
string is described by
\begin{mathletters}\label{ad-str-sol}
\begin{eqnarray}
(X^{\mu})&=&(t(\lambda,\sigma),x(\lambda,\sigma),0,0)  \\
\epsilon&=&\epsilon_0(\lambda)  \\
\tau&=&\tau(\epsilon_0(\lambda))\equiv\tau_0(\lambda)  \\
(u^a)&=&(e^{-H\lambda},0).
\end{eqnarray}
\end{mathletters}%
where $\epsilon_0(\lambda)$ is the solution of
Eq.~(\ref{eq-epsilon_0}), and $t(\lambda,\sigma)$ and
$x(\lambda,\sigma)$ are given in Eqs.~(\ref{transformation}).\par
Let us first discuss longitudinal wiggles. In that case, the perturbed
string is described by
\begin{mathletters}\label{long-pert-string}
\begin{eqnarray}
(X^{\mu})&=&(t(\lambda,\sigma),x(\lambda,\sigma),0,0)  \\
\epsilon&=&\epsilon_0(\lambda)+\epsilon_1(\lambda,\sigma) \\
\tau&=&\tau(\epsilon)=\tau_0(\lambda)
 -v_{\scriptscriptstyle L,0}^2(\lambda)\epsilon_1(\lambda,\sigma)
 +{\cal O}(\epsilon_1^2) \\
(u^a)&=&{e^{-H\lambda}\over \sqrt{1-\beta^2}}(1,\beta),
\end{eqnarray}
\end{mathletters}%
with $t(\lambda,\sigma)$ and $x(\lambda,\sigma)$ still given by
Eqs.~(\ref{transformation}), and
$v_{\scriptscriptstyle L,0}^2(\lambda)=
    -{d\tau\over d\epsilon}(\epsilon_0(\lambda))$. The
    stress-energy-momentum tensor is
\begin{equation}
(t^{ab})=e^{-2H\lambda}
\pmatrix{(\epsilon-\tau)\gamma^2+\tau&(\epsilon-\tau)\gamma^2\beta\cr
(\epsilon-\tau)\gamma^2\beta&(\epsilon-\tau)\gamma^2\beta^2-\tau}.
\end{equation}
The equations of motion may be shown to be equivalent to
\begin{mathletters}\label{eqmo-comp}
\begin{eqnarray}
\partial_{\lambda}[(\epsilon-\tau)\gamma^2+\tau]
 +\partial_{\sigma}[(\epsilon-\tau)\gamma^2\beta]
 +H(\epsilon-\tau)\gamma^2(1+\beta^2)&=&0  \\
\partial_{\lambda}[(\epsilon-\tau)\gamma^2\beta]
 +\partial_{\sigma}[(\epsilon-\tau)\gamma^2\beta^2-\tau]
 +2H(\epsilon-\tau)\gamma^2\beta&=&0  .
\end{eqnarray}
\end{mathletters}%
{}From Eqs.~(\ref{eqmo-comp}), one readily obtains the equations of
motion first order in $\beta$ and
$\delta\equiv{\epsilon_1\over\epsilon_0}$:
\begin{mathletters}\label{eqmo-beta-delta}
\begin{eqnarray}
\partial_{\lambda}\delta+
H(v_{\scriptscriptstyle L,0}^2+v_{\scriptscriptstyle T,0}^2)\delta+
  (1-v_{\scriptscriptstyle T,0}^2)\partial_{\sigma}\beta&=&0  \\
(1-v_{\scriptscriptstyle T,0}^2)\partial_{\lambda}\beta+
  H(1-v_{\scriptscriptstyle T,0}^2)
  (1-v_{\scriptscriptstyle L,0}^2)\beta+
  v_{\scriptscriptstyle L,0}^2\partial_{\sigma}\delta&=&0.
\end{eqnarray}
\end{mathletters}%
Since the coefficients in Eqs.~(\ref{eqmo-beta-delta}) do not depend
upon $\sigma$, we may write
\begin{equation}
\beta(\lambda,\sigma)=\beta(\lambda)e^{ik\sigma},
\quad\delta(\lambda,\sigma)=\delta(\lambda)e^{ik\sigma},
\label{fourier-comp}
\end{equation}
where $k$ is the wavevector in the comoving coordinates.
We are interested in the regime where the wiggle is being stretched
adiabatically. This requires $H\ll k$. Therefore we drop terms of
order $H^2$, $H{d v_{\scriptscriptstyle L,0}\over d\lambda}$,
{\it  etc.}, as we rearrange Eqs.~(\ref{eqmo-beta-delta}) to
separate $\delta$ and $\beta$:
\begin{mathletters}\label{eqmo-separate}
\begin{eqnarray}
\ddot{\delta}+H(1+2v_{\scriptscriptstyle T,0}^2+
v_{\scriptscriptstyle L,0}^2)\dot{\delta} +
v_{\scriptscriptstyle L,0}^2 k^2\delta&=&0 \\
\ddot{\beta}+
[{d\over d\lambda}\ln({1-v_{\scriptscriptstyle T,0}^2
\over v_{\scriptscriptstyle L,0}^2})
+H(1+v_{\scriptscriptstyle T,0}^2)]\dot{\beta}+
v_{\scriptscriptstyle L,0}^2 k^2\beta&=&0.
\end{eqnarray}
\end{mathletters}%
Because the coefficients in Eqs.~(\ref{eqmo-separate}) are slowly
varying functions of $\lambda$, we may solve these equations  by
the method of adiabatic invariants. This yields
\begin{mathletters}\label{sol-delta-beta}
\begin{eqnarray}
\delta(\lambda,\sigma)&=&C e^{-H\lambda}
 \Big[{1-v_{\scriptscriptstyle T,0}^2(\lambda)\over
 \epsilon_0(\lambda)v_{\scriptscriptstyle L,0}(\lambda)}\Big]^{1/2}
 e^{ik[\sigma\pm\int^{\lambda}
  d\lambda'v_{\scriptscriptstyle L,0}(\lambda')]} \label{sdba}\\
\beta(\lambda,\sigma)&=&-Ce^{-H\lambda}
 \Big[{v_{\scriptscriptstyle L,0}(\lambda)\over
 (1-v_{\scriptscriptstyle T,0}^2(\lambda))\epsilon_0(\lambda)}\Big]^{1/2}
 e^{ik[\sigma\pm\int^{\lambda}
  d\lambda'v_{\scriptscriptstyle L,0}(\lambda')]},\label{sdbb}
\end{eqnarray}
\end{mathletters}%
where $C$ is a constant. These equations tell us how the amplitude and
the phase of a longitudinal wiggle change when the string is slowly
stretched.\par
As was mentioned in section I and III, we modified in this paper the
definition of what it means to add a wiggle to a string, from what it was
in ref.~\cite{HKS92}.
We are finally in a position to explain what prompted the modification.
With the definition used in ref.~\cite{HKS92}, one
has, instead of Eq.~(\ref{epsi2}),
\begin{equation}
\langle\epsilon_2\rangle=(\epsilon_0-\tau_0)\langle\beta^2\rangle
\label{epsi2ref6}
\end{equation}
for longitudinal wiggles. Hence there are in this case two equal
contributions, $\langle\epsilon_2\rangle$ and
$(\epsilon_0-\tau_0)\langle\beta^2\rangle$, to the renormalization of
$\epsilon_0$ (see Eq.~(\ref{sec-emtensor})). However, when this string
is stretched and the analysis presented above is applied to it, one finds
that $\langle\epsilon_2\rangle$ and
$(\epsilon_0-\tau_0)\langle\beta^2\rangle$ do not remain equal during
the stretching. Their rates of decrease are different. The reason for
this phenomenon is that what was called ``adding a longitudinal wiggle
to a string'' in ref.~\cite{HKS92} actually has two different
components. The first component is a homogeneous modification of the
original string, and the second component is adding a wiggle. The two
components behave differently under homogeneous stretching. \par
Next, let us consider transverse wiggles.
The wiggly string is described by
\begin{mathletters}\label{trans-pert-string}
\begin{eqnarray}
(X^{\mu})&=&(t(\lambda,\sigma),x(\lambda,\sigma),y(\lambda,\sigma),0)  \\
\epsilon&=&\epsilon_0(\lambda) +\epsilon_1(\lambda,\sigma)  \\
\tau&=&\tau(\epsilon)=\tau_0(\lambda)
  -v_{\scriptscriptstyle L,0}^2\epsilon_1(\lambda,\sigma)
  +{\cal O}(\epsilon_1^2) \\
(h_{ab})&=&e^{2H\lambda}\pmatrix{1&0\cr 0&1}
  -\pmatrix{{\dot{y}}^2&\dot{y}y'\cr \dot{y}y'&{y'}^2} \\
(u^a)&=&{(1,\beta)\over
\sqrt{e^{2H\lambda}(1-\beta^2)-(\dot{y}+\beta{y'})^2}}.
\end{eqnarray}
\end{mathletters}%
The linearized equation of motion for y, which follows from the
$\mu=2$ component of Eqs.~(\ref{4eqmo}), is
\begin{equation}
\partial_{\lambda}[\epsilon_0(\lambda) \partial_{\lambda}\,y]
-\tau_0(\lambda)\partial_{\sigma}^2\,y=0\, .\label{eqmo-trans-y}
\end{equation}
Again we set $y(\lambda,\sigma)=y(\lambda)\,e^{ik\sigma}$
and use the method of adiabatic invariants to find
\begin{equation}
y(\lambda,\sigma)={C'\over [\epsilon(\lambda)\tau(\lambda)]^{1/4}}
 e^{ik[\sigma\pm\int^{\lambda}
 d\lambda'v_{\scriptscriptstyle T,0}(\lambda')]}. \label{sol-y}
\end{equation}
Eq.~(\ref{sol-y}) tells us how the amplitude and the phase of a
transverse wiggle change when the string is slowly stretched.
\section{Renormalization of the equation of state due to wiggliness}
The equation of state function
$\tau(\epsilon)$ is equivalent to the infinite set of parameters:
$\epsilon$, $\tau$,
${d\tau\over d\epsilon}\equiv -v_{\scriptscriptstyle L}^2$,
${d^2\tau\over d\epsilon^2}$, ${d^3\tau\over d\epsilon^3}$, $\ldots\,$.
In ref.~\cite{HKS92} and in section III of this paper, the renormalization
of the first two of these parameters,
$\epsilon$ and $\tau$, which results from averaging out small scale
wiggles on the string was obtained to lowest order in the amount of
wiggliness. It has the form:
\begin{mathletters}\label{renorm_skeleton-again}
\begin{eqnarray}
\bar\epsilon&=&\epsilon +
C_{\epsilon}(\epsilon,\tau,v_{\scriptscriptstyle L},\ldots\,)
\langle v^2\rangle \\
\bar\tau&=&\tau +
C_{\tau}(\epsilon,\tau,v_{\scriptscriptstyle L},\ldots\,)
\langle v^2\rangle
\end{eqnarray}
\end{mathletters}%
where $\langle v^2\rangle$ is the average (velocity)$^2$ associated
with the wiggle.
The coefficients $C_\epsilon$ and
$C_\tau$ are given in Eqs.~(\ref{long-renorm}) and
(\ref{trans-renorm})  for longitudinal and transverse wiggles
respectively. In the previous section, we derived the response of the
wiggles to a slow stretching of the string. We now use these results
to derive, in principle, the renormalization of all the other parameters
(${d^n\tau\over d\epsilon^n}$, $n=1,2,3,\ldots\,$) that characterize the
equation of state. \par
We wish to determine the coefficients  $C_n$:
\begin{equation}
{d^n\bar\tau\over d\bar\epsilon^n}=
 {d^n\tau\over d\epsilon^n}+C_n\langle v^2\rangle+
 {\cal O}(v^4) \label{eq-C_n}
\end{equation}
for $n=1,2,3,\ldots\,$. What we will  succeed in doing is to derive a
recursion formula for these $C_n$. Let us call
${1\over H}{dX\over d\lambda}$ the rate of change of quantity $X$
under slow stretching. The notation ${1\over H}{d\, \over d\lambda}$
is taken from the previous section. We have
\begin{mathletters}\label{recurs-eq-param}
\begin{eqnarray}
&&{1\over H}{d\epsilon\over d\lambda}=-(\epsilon-\tau) \label{repa}\\
&&{1\over H}{d\,\over d\lambda}{d^n\tau\over d\epsilon^n}=
 -(\epsilon-\tau){d^{n+1}\tau\over d\epsilon^{n+1}}  \label{repb}\\
&&{1\over H}{d\,\over d\lambda}C_n=
 -(\epsilon-\tau){dC_n\over d\epsilon}. \label{repc}
\end{eqnarray}
\end{mathletters}%
Eq.~(\ref{repa}) is the same as Eq.~(\ref{eq-epsilon_0}).
Eqs.~(\ref{repb}) and (\ref{repc}) follow from the fact that,
through the equation of state, $\tau$, ${d^n\tau\over d\epsilon^n}$,
$C_\epsilon$, $C_\tau$, $C_n$ are functions of $\epsilon$ only. The
crucial information gained in the previous section is the rate of
change of $\langle v^2\rangle$ under slow stretching.
Let us define coefficients $D_v$ by
\begin{equation}
{1\over H}{d\,\over d\lambda}\langle v^2\rangle=
 -D_v \langle v^2\rangle. \label{recurs-eq-beta}
\end{equation}
Expressions for the coefficients $D_v$
for both transverse and longitudinal wiggles may be obtained from
Eqs.~(\ref{sdbb}) and (\ref{sol-y}). They are given below.
Suffice it to say for the moment that the coefficients
$D_v$, like the
coefficients $C_n$, are functionals of the equation of state only.
Combining Eqs.~(\ref{renorm_skeleton-again}),
(\ref{eq-C_n}), (\ref{recurs-eq-param}) and (\ref{recurs-eq-beta}),
we have
\begin{equation}
{d^{n+1}\bar\tau\over d\bar\epsilon^{n+1}}=
 {{1\over H}{d\,\over d\lambda}{d^n\bar\tau\over d\bar\epsilon^n}
   \over {1\over H}{d\,\over d\lambda}\bar\epsilon}=
 {{d^{n+1}\tau\over d\epsilon^{n+1}}+\langle v^2\rangle
  [{dC_n\over d\epsilon}+{C_nD_v\over \epsilon-\tau}]\over
  1+\langle v^2\rangle
  [{dC_{\epsilon}\over d\epsilon}+{C_{\epsilon}D_v\over \epsilon-\tau}]}.
\label{master-recurs-eq}
\end{equation}
Hence, the recursion formula:
\begin{equation}
C_{n+1}={dC_n\over d\epsilon}+{C_nD_v\over \epsilon-\tau}
 -{d^{n+1}\tau\over d\epsilon^{n+1}}
  \Big[{dC_{\epsilon}\over d\epsilon}+
    {C_{\epsilon}D_v\over \epsilon-\tau}\Big],
\label{master-recurs-eq-C_n}
\end{equation}
which allows one, in principle, to compute all the coefficients $C_n$
successively starting with $C_\tau\equiv C_0$. However, for a general
equation of state, the coefficients $C_n$ quickly
become very complicated.\par
For transverse wiggles, using Eqs.~(\ref{sol-y}), (\ref{repa})
and (\ref{repb}) and
the fact that ${1\over H}{d\,\over d\lambda}\ln k_{\rm phys}=-1$, we obtain
\begin{eqnarray}
{1\over H}{d\,\over d\lambda}\langle\dot{y}^2\rangle&=&
  {1\over H}{d\,\over d\lambda}
   [k^2_{\rm phys}v_{\scriptscriptstyle T}^2
    \langle y^2\rangle] \nonumber \\
&=&\langle\dot{y}^2\rangle
  [-2 + {1\over 2}(1-{\tau\over\epsilon})
(3+v_{\scriptscriptstyle L}^2)] \label{trans-lambda-dep}
\end{eqnarray}
where $k_{\rm phys}$ is the physical wavevector as opposed to the
comoving wavevector $k$ introduced in Eq.~(\ref{fourier-comp}).
Thus for transverse wiggles, from Eqs.~(\ref{trans-renorm}) and
(\ref{trans-lambda-dep}), we have
\begin{mathletters}\label{trans-sol-coeffs}
\begin{eqnarray}
\langle v^2\rangle&=&\langle\dot{y}^2\rangle+
      \langle\dot{z}^2\rangle  \label{tsca} \\
C_\epsilon&=&\epsilon \label{tscb} \\
C_\tau&=&C_0=-{1\over 2}
 [\tau+\epsilon-\epsilon(1-{\epsilon\over\tau}){d\tau\over d\epsilon}]
  \label{tscc} \\
D_v&=&+2-{1\over 2}
  (1-{\tau\over\epsilon})(3-{d\tau\over d\epsilon}). \label{tscd}
\end{eqnarray}
\end{mathletters}%
These expressions may be fed into Eq.~(\ref{master-recurs-eq-C_n}) to
obtain  the coefficients $C_n$ for transverse wiggles. The first
one is
\begin{eqnarray}
C_1&=&{1\over 4}+{3\over 4}{\tau\over\epsilon}
  -{\tau+\epsilon\over\epsilon-\tau}
  -{d\tau\over d\epsilon}\Big[{1\over 2}
  +{1\over 4}{\epsilon\over\tau}+{1\over 4}{\tau\over\epsilon}
  +{(\epsilon+\tau)\epsilon\over\tau(\epsilon-\tau)}\Big]\nonumber \\
&&\quad +{1\over 2}{d^2\tau\over d\epsilon^2}
  \epsilon(1-{\tau\over\epsilon})
  +{1\over 2}\Big({d\tau\over d\epsilon}\Big)^2
   \Big[{\epsilon^2\over\tau^2}-{1\over 2}(1+{\epsilon\over\tau})\Big].
\label{trans-sol-C_1}
\end{eqnarray}
\par
For longitudinal wiggles, using Eqs.~(\ref{sdbb}) and
(\ref{recurs-eq-param}), we obtain
\begin{eqnarray}
{1\over H}{d\,\over d\lambda}\langle\beta^2\rangle&=&
  {1\over H}{d\,\over d\lambda}
   \Big[C^2 e^{-2H\lambda}
   {v_{\scriptscriptstyle L}\over\epsilon(\lambda)-\tau(\lambda)}
   \Big]   \nonumber \\
&=&\langle\beta^2\rangle
  \Big[-1+v_{\scriptscriptstyle L}^2-(\epsilon-\tau)
   {d\ln{v_{\scriptscriptstyle L}}\over d\epsilon}\Big] .
   \label{long-lambda-dep}
\end{eqnarray}
{}From Eqs.~(\ref{sec-emtensor}) and (\ref{long-lambda-dep}), we have then
\begin{mathletters}\label{long-sol-coeffs}
\begin{eqnarray}
\langle v^2\rangle&=&\langle\beta^2\rangle \label{lsca}\\
C_\epsilon&=&\epsilon-\tau \label{lscb} \\
C_\tau&=&-(\epsilon-\tau)\Big[1+ (\epsilon-\tau)
   {d\ln{v_{\scriptscriptstyle L}}\over d\epsilon}\Big]
    \label{lscc} \\
D_v&=&1-v_{\scriptscriptstyle L}^2 + (\epsilon-\tau)
   {d\ln{v_{\scriptscriptstyle L}}\over d\epsilon}. \label{lscd}
\end{eqnarray}
\end{mathletters}%
These expressions may be fed into Eq.~(\ref{master-recurs-eq-C_n}) to
obtain the coefficients $C_n$ for longitudinal wiggles. The first one
is
\begin{eqnarray}
C_1&=&-2-2{d\tau\over d\epsilon}-4(\epsilon-\tau)
  {d\ln{v_{\scriptscriptstyle L}}\over d\epsilon} \nonumber \\
&&\quad -(\epsilon-\tau)^2
 \Big({d\ln{v_{\scriptscriptstyle L}}\over d\epsilon}\Big)^2
 -(\epsilon-\tau)^2
 {d^2\ln{v_{\scriptscriptstyle L}}\over d\epsilon^2}.\label{long-sol-C_1}
\end{eqnarray}
\section{Wiggly Nambu-Goto Strings}
In this section, we apply our formalism to the case of
gauge strings or any other strings which at the microscopic
level obey the Nambu-Goto (NG) equation of state $\epsilon=\tau=\mu$.
In this case,
\begin{mathletters}\label{sol-NG-C}
\begin{eqnarray}
C_{\epsilon{\scriptscriptstyle T}}&=&\mu, \quad
C_{\tau{\scriptscriptstyle T}}=-\mu  \label{sNCa}\\
D_{v{\scriptscriptstyle T}}&=&2 \label{sNCb}\\
C_{\epsilon{\scriptscriptstyle L}}&=&
 C_{\tau{\scriptscriptstyle L}}=0  ,
\label{sNCc}
\end{eqnarray}
\end{mathletters}%
where the subscripts $T$ and $L$ stand for transverse and longitudinal
respectively.
Eq.~(\ref{sNCc}) reflects the fact that Nambu-Goto strings can not
have longitudinal oscillations. They may have transverse oscillations
and Eq.~(\ref{sNCa}) tells us that the string parameters
characterizing  the renormalized wiggly Nambu-Goto (RWNG) string are
\begin{mathletters}\label{sol-RWNG}
\begin{eqnarray}
\bar\epsilon&=&
\mu(1+\langle\dot{y}^2\rangle+\langle\dot{z}^2\rangle) \label{sRa}\\
\bar\tau&=&
\mu(1-\langle\dot{y}^2\rangle-\langle\dot{z}^2\rangle) \label{sRb}
\end{eqnarray}
\end{mathletters}%
to lowest order in $\langle\dot{y}^2\rangle$ and
$\langle\dot{z}^2\rangle$. From Eq.~(\ref{sNCb}), or   more directly
from  Eq.~(\ref{sol-y}) with $\epsilon_0=\tau_0=\mu$, we learn that
if a wiggly Nambu-Goto string  is stretched, the amplitude of the wiggle does
not change. Vilenkin\cite{vilenkin85}
had already established this fact in the case of
a wiggly Nambu-Goto string that is being stretched by the expansion
of the universe.  A simple way to derive this result is as follows.
Under homogeneous stretching, using Eqs.~(\ref{sol-RWNG}), we have:
\begin{eqnarray}
{1\over H}{d\bar\epsilon\over d\lambda}&=&
 -(\bar\epsilon-\bar\tau)=
 -2\mu(\langle\dot{y}^2\rangle+\langle\dot{z}^2\rangle) \nonumber \\
&=&{\mu\over H}{d\, \over d\lambda}
(\langle\dot{y}^2\rangle+\langle\dot{z}^2\rangle) .
\end{eqnarray}
This implies
\begin{eqnarray}
&&{1\over H}{d\, \over d\lambda}
\ln(\langle\dot{y}^2\rangle+\langle\dot{z}^2\rangle) \nonumber \\
&=&{1\over H}{d\, \over d\lambda}
\ln[k^2_{\rm phys}(\langle y^2\rangle+\langle z^2\rangle)]=-2 ,
\end{eqnarray}
and hence
\begin{equation}
{1\over H}{d\, \over d\lambda}
(\langle y^2\rangle+\langle z^2\rangle)=0
\end{equation}
since ${1\over H}{d\, \over d\lambda}\ln k_{\rm phys}=-1$. While
the string is being stretched, $\bar\epsilon$ decreases and $\bar\tau$
increases in such a way as to keep $\bar\epsilon\bar\tau=\mu^2$. This
is the equation of state of the RWNG string to the order to which we
have calculated. \par
Now we add wiggles to the RWNG string. We may do that if
the wavelength of the new wiggles are long compared to the wavelength
of the wiggles that have already been averaged over. The equation of
state that must be fed into Eqs.~(\ref{long-renorm}) and
(\ref{trans-renorm}) to obtain the renormalized values of $\epsilon$
and $\tau$ which result from averaging out wiggles on the RWNG string
is $\epsilon\tau=\mu^2$. But  it now turns out,   remarkably, that the
equation of state $\epsilon\tau=\mu^2$ is a fixed point of the
renormalization group equations (\ref{RGE}). Indeed one may readily
verify that
\begin{equation}
{d\, \over d\ln k}(\epsilon\tau)=0
\end{equation}
for all $W_{\scriptscriptstyle T}(k)$ and $W_{\scriptscriptstyle L}(k)$
if the equation of state is $\epsilon\tau=\mbox{constant}$. This means
that $\epsilon\tau=\mu^2$ is the equation of state of renormalized
wiggly Nambu-Goto strings no matter how many wiggles are added to the
string providing only that $W_{\scriptscriptstyle T}(k)$ and
$W_{\scriptscriptstyle L}(k)$, as functions of wavevector $k$, are
everywhere sufficiently small. The equation of state
$\epsilon\tau=\mu^2$ for renormalized   wiggly Nambu-Goto strings was
conjectured by Carter\cite{carter90} and Vilenkin\cite{vilenkin90}.
In ref.~\cite{HKS92}, using the argument
just given, it was shown that $\epsilon\tau=\mu^2$ is indeed the
equation of state of renormalized wiggly Nambu-Goto strings to lowest
non-trivial order in the amount of wiggliness. But it was also
explicitly shown in ref.~\cite{HKS92} that in higher orders there are
deviations from $\epsilon\tau=\mu^2$.\par
Fig.\ II provides a summary. The short-dashed lines represent
the paths in $(\epsilon,\tau)$ space which are followed when  a string
is homogeneously stretched. The short-dashed lines are therefore given by
the equation of state. The long-dashed lines represent the paths in
$(\epsilon,\tau)$ space which are followed when one renormalizes
$\epsilon$ and $\tau$ to average out small scale wiggles. The long-dashed
lines are therefore given by integrating the renormalization group
equations (\ref{RGE}). The central aim of this paper is to derive the
modification of the equation of state that results from averaging out
small scale wiggles, {\it i.e.}, to determine how the short-dashed lines
are related to one another when one moves along the long-dashed lines. Our
results in  this regard are stated in sections V and VI. The equation
of state $\epsilon\tau=\mu^2$ is shown by the solid line in Fig.~II.
It has the following special property: if at any point on the solid
line  the short-dashed line is parallel to the solid line, then the
long-dashed
line is also parallel to the solid line there. As a consequence of
this and the fact that $\epsilon\tau=\mu^2$ for a wiggly Nambu-Goto
string near $\epsilon=\tau=\mu$, when  adding wiggles to such a string
and averaging over them, one remains forever on the
$\epsilon\tau=\mu^2$ line, to lowest order in perturbation theory.\par
Finally, let us apply the results of section V to the RWNG string.
Using the equation of state $\epsilon\tau=\mu^2$, Eq.~(\ref{sol-y})
implies that the amplitude of transverse wiggles on a RWNG string
stays constant when the string is slowly stretched, just as  is the
case for the Nambu-Goto string. In addition, Eq.~(\ref{sdbb}) tells us
that the $\beta$-amplitude of a longitudinal wiggle on a RWNG string
is constant when the string is slowly stretched.
\acknowledgements
We are grateful to Jooyoo Hong for useful discussions. This work was
supported in part by the U.~S.~Department of Energy under contract No.\
DE-FG05-86ER40272.

\begin{figure}
\caption{Comoving Orthogonal Coordinates parametrizing a homogeneously
stretching string: the radial lines which emerge
from $(t,x)=(-{1\over H},0)$ are equi-$\sigma$,
while the hyperbolas are equi-$\lambda$. Note that the homogeneously
stretching string may be defined only inside the future light
cone of $(t,x)=(-{1\over H},0)$.}
\label{fig1}
\end{figure}
\begin{figure}
\caption{Behaviour of string parameters $\epsilon$ and $\tau$
under adiabatic stretching and under renormalization to average out
small scale wiggles. The point of coordinates $(\epsilon,\tau)$
moves along a short-dashed line when the string is adiabatically stretching,
whereas it moves along a long-dashed line when $\epsilon$ and $\tau$ are
redefined to average out small scale wiggles. The solid line
represents RWNG strings for which $\epsilon\tau=\mu^2$,
while the blob at the end corresponds to a NG string.}
\label{fig2}
\end{figure}
\end{document}